\documentclass[useAMS,usenatbib]{mn2e}
\usepackage{journals}
\usepackage{times}
\usepackage{float}

\newcommand{\sub}[1]{_{\mathrm{#1}}}

\newcommand{\ergscm}[1]{$10^{#1}$~erg~cm$^{-2}$~s$^{-1}$}

\newcommand{\pcu}{c~s$^{-1}$PCU$^{-1}$}

\newcommand{\cont}{4U 1728-34}
\newcommand{\grs}{GRS 1915+105}
\newcommand{\igr}{IGR J17091-3624}

\usepackage{url}
\usepackage{amssymb}
\usepackage{graphicx}
\usepackage{textcomp}

\title[\grs\ variability patterns in the RB]{Discovery of \grs\ variability patterns in the Rapid Burster} 
\author[Bagnoli \& in 't Zand]{T. Bagnoli$^{1,2}$\thanks{E-mail: t.bagnoli@sron.nl},
and J.J.M. in 't Zand$^{1}$
\\
$^{1}$SRON Netherlands Institute for Space Research,
Sorbonnelaan 2, 3584 CA Utrecht, The Netherlands\\
$^{2}$Astronomical Institute ``Anton Pannekoek'', University of Amsterdam,
Postbus 94249, 1090 GE Amsterdam, The Netherlands
\\
\\
\\
\textup{Accepted for publication in MNRAS letters}
}

\begin{document}

\date{}

\pagerange{\pageref{firstpage}--\pageref{lastpage}} \pubyear{}

\maketitle

\label{firstpage}

\begin{abstract}

We report the discovery of two new types of variability
in the neutron star low-mass X-ray binary MXB 1730-335 (the 'Rapid Burster').
In one observation in 1999, it exhibits a large-amplitude quasi-periodic oscillation
with a period of about 7 min.
In another observation in 2008, it exhibits two 4-min long 75 per cent deep dips 44 min apart.
These two kinds of variability are very similar to the so-called $\rho$ or 'heartbeat'
variability and the $\theta$ variability, respectively,
seen in the black hole low-mass X-ray binaries \grs\ and \igr.
This shows that these types of behavior are unrelated
to a black hole nature of the accretor.
Our findings also show that these kinds of behaviour
need not take place at near-Eddington accretion rates.
We speculate that they may rather be related to the presence
of a relatively wide orbit with an orbital period in excess of a few days
and about the relation between these instabilities and
the type II bursts.

\end{abstract}


\section{Introduction}
\label{sec:intro}

The Rapid Burster (RB, or MXB 1730-335) was discovered by
\citet{1976ApJ...207L..95L}
as a source of few-seconds long X-ray bursts with
recurrence times as small as 6 s, much faster than similar bursts in
other sources
with recurrence times of an hour or longer.
The RB itself was later resolved to exhibit both
kinds of X-ray bursts which were from then on called (slow)
thermonuclear type I and (rapid) accretion-powered type II X-ray
bursts \citep{1978Natur.271..630H}.
Up to this date, this dual character of the RB
remains unique. The RB is located in the globular cluster Liller 1,
which yields an independent distance estimate of
7.9$\pm$0.9 kpc \citep{2010MNRAS.402.1729V}.

Type I X-ray bursts have so far been seen from about 100 low-mass X-ray
binaries in our galaxy. The accretor is a neutron star (NS), on which
H and/or He is piled up from the atmosphere of the
Roche-lobe overflowing companion star. After a certain time, typically a
few hours, ignition conditions are reached at the bottom of the pile
and a thermonuclear runaway burns the hydrogen and helium within a
fraction of a second. Subsequently, the photosphere heats up to a
few tens of MK and cools off, giving rise to a softening X-ray burst
of duration $\sim$1~min (see reviews by \citealt*{1993SSRv...62..223L} and
\citealt{2006csxs.book..113S}).

Type II bursts have so far only been identified in two low-mass
X-ray binaries: the RB and the Bursting Pulsar (BP). The BP
\citep{1996Natur.379..799K} is a transient accretion-powered X-ray pulsar with a
period of 0.467~s and a magnetic dipole field strength of a few times
10$^{10}$~G, and a probably low-mass giant companion star in a 11.8~d
orbit \citep{1996Natur.381..291F,2014ApJ...796L...9D}.

The RB is less well defined than the BP.
Neither the NS spin frequency nor the dipolar magnetic field
are known because of the lack of X-ray pulsations.
This could either be due to a magnetic field smaller than in the
BP (indeed the majority of bursters do not show
persistent pulsations), or because of a close alignment of either
the observer's line of sight or the magnetic dipolar component
with the rotation axis.
The orbital period is also not known.
The RB is a transient that goes into outbursts
every 100--200~days. The accretion
rate increases to quite high values, close to the Eddington limit
\citep{2013MNRAS.431.1947B}, which is rare among LMXBs.

The RB was abundantly observed with the Proportional Counter
Array (PCA) on the \textit{Rossi X-ray Timing Explorer} (\textit{RXTE}).
A total exposure time of
2.4~Ms was obtained in 17 years of observations,
and the RB was active during 1.4~Ms of these
\citep{2015Ba}. The large exposure time and large
sensitivity of the PCA provide excellent opportunity to detect rare
behavior in the RB.
In this letter we present the detection
of two kinds of peculiar behaviour of the RB,
and speculate on their implications.

\begin{figure}
\includegraphics[width=\columnwidth,angle=0]{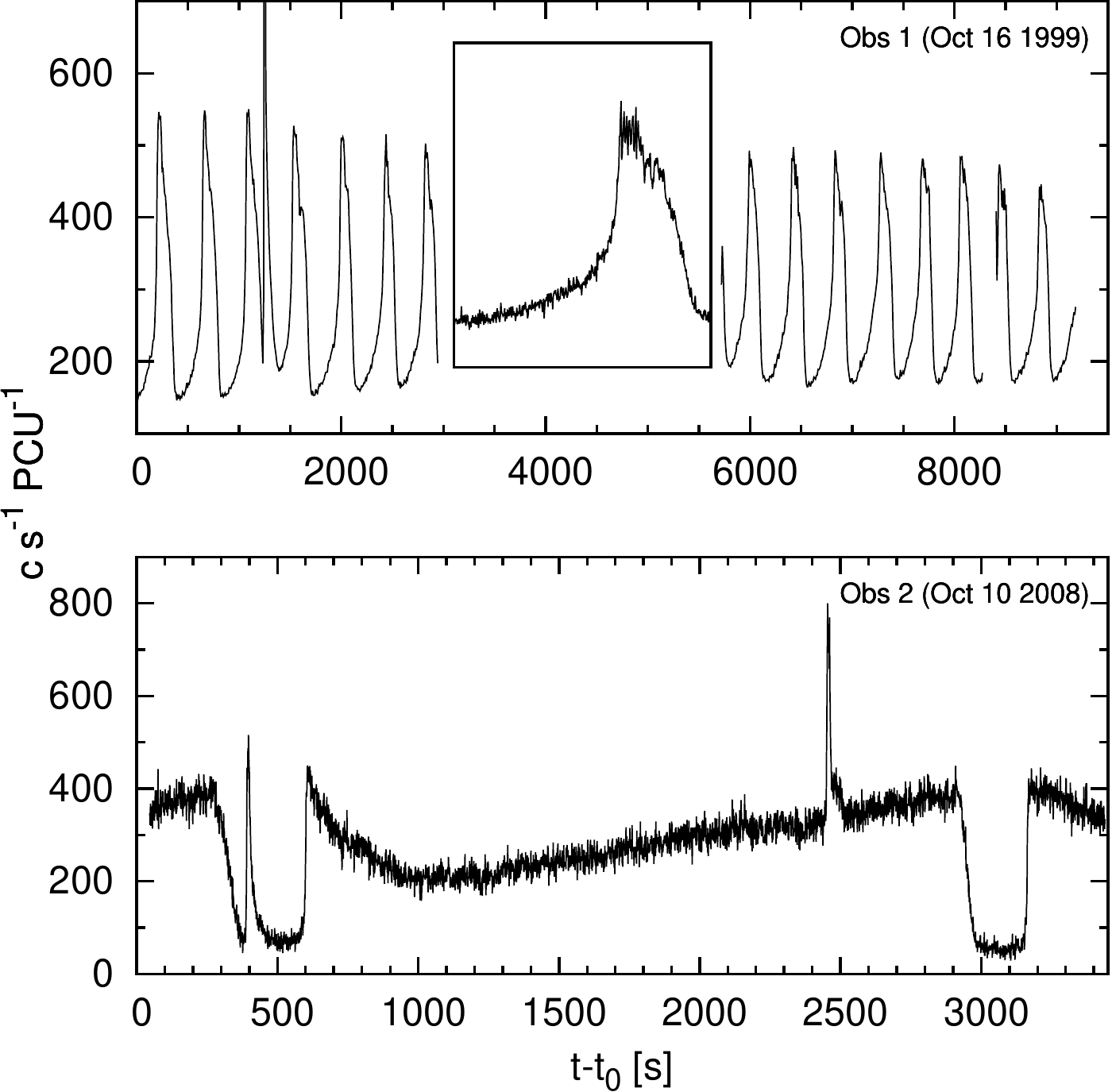}
\caption{Light curves of the two observations
		from the full bandpass at 1~s resolution.
The intensity was corrected for the background
and collimator response to the RB,
and for the contamination by \cont\
during the first part of the first observation
(see Fig.~\ref{fig:hid}).
		The inset in the top panel covers a data gap and
		shows a complete cycle (450~s) of the instability.
\label{fig:lc}}
\end{figure}

\section{Observations}
\label{sec:obs}

The PCA consists of five proportional counter units (PCUs), each with
a photon-collecting area of 1600 cm$^2$ \citep{2006ApJS..163..401J}, between 2--60~keV
with a spectral resolution of 20 per cent full-width at half
maximum (FWHM) at 6 keV.
The two observations that are presented here date to Oct. 16, 1999
(ObsID 40433-01-04-00R and -01R, exposure time 2.8 and 3.4~ks with a
data gap of 3.8~ks)
and Oct. 10, 2008
(ObsID 92026-01-20-02, exposure time 3.2~ks)
The RB was on-axis during the first observation except for the
last 0.8~ks when the nearby bright \cont\ was slewed out of the
field of view, thereby putting the RB at an off-axis angle of 0.56~deg
and avoiding contaminating signal from \cont.
The pointing of the second observation was
identical to the last 0.8~ks of the first observation,
excluding the signal from \cont.

Figure ~\ref{fig:lc} (upper panel)
shows the light curve of the first observation.
A strong quasi-periodic oscillation (QPO) is apparent with a
period varying from $\sim350$ to $\sim450$~s and a
peak-to-peak amplitude of roughly 70 per cent
(from $\approx600$ to $\approx200$~\pcu).
We observed 15 such flares. 
The last two flares occurred during a pointing
uncontaminated by 4U 1728-34.
This behavior has not been seen from the RB in
any other \textit{RXTE} observation or in any other published observation.
Near the 1300~s mark, a burst takes place,
lasting about 100~s and significantly sub-Eddington.
The characteristics of this burst are consistent with being of type I:
it shows a decreasing spectral hardness and lacks ringing during
the decay so typical of type II bursts of this duration
\citep{2015Ba}.
The type-I burst is identified to originate from the RB,
because its peak flux, duration and lack of photospheric
expansion are consistent with other bursts from the RB
and inconsistent with all other bursts from 4U 1728-34
\citep[for more details on the burst identification, see ][]{2013MNRAS.431.1947B}.
The light curve of the second observation is plotted
in Fig.~\ref{fig:lc} (lower panel). It
shows a gradual change in intensity with two dips by
about 75 per cent in intensity lasting 260 and 220~s,
respectively (from halfway ingress to halfway egress).
The ingresses are about 100 and 80~s long,
while the egresses last a shorter 40 and 20~s.
The persistent emission outside dips peaks at about 400~\pcu
right after the first dip,
falls to 200~\pcu in 400~s,
and then peaks again at the roughly the same maximum value
right before the second one.
The entire cycle (ingress, dip, egress, maximum, 
decreasing and increasing back to maximum)
lasts about 2650~s.
The second dip is on average about 50~\pcu.
Judging from the portions of the lightcurve before the first
dip and after the second, the behaviour seems to be roughly cyclical.
Also, there are two bursts, one of them in a dip.
 These are also of type I, following the above-mentioned arguments.
They must be from the RB
because \cont\ was outside the field of view.
Both have a net peak flux of 460~\pcu,
calculated with respect to the persistent emission flux right before.
The type I burst recurrence time is 2.1~ks.

\section{Data analysis}
\label{sec:analysis}

\begin{figure}
\includegraphics[width=\columnwidth]{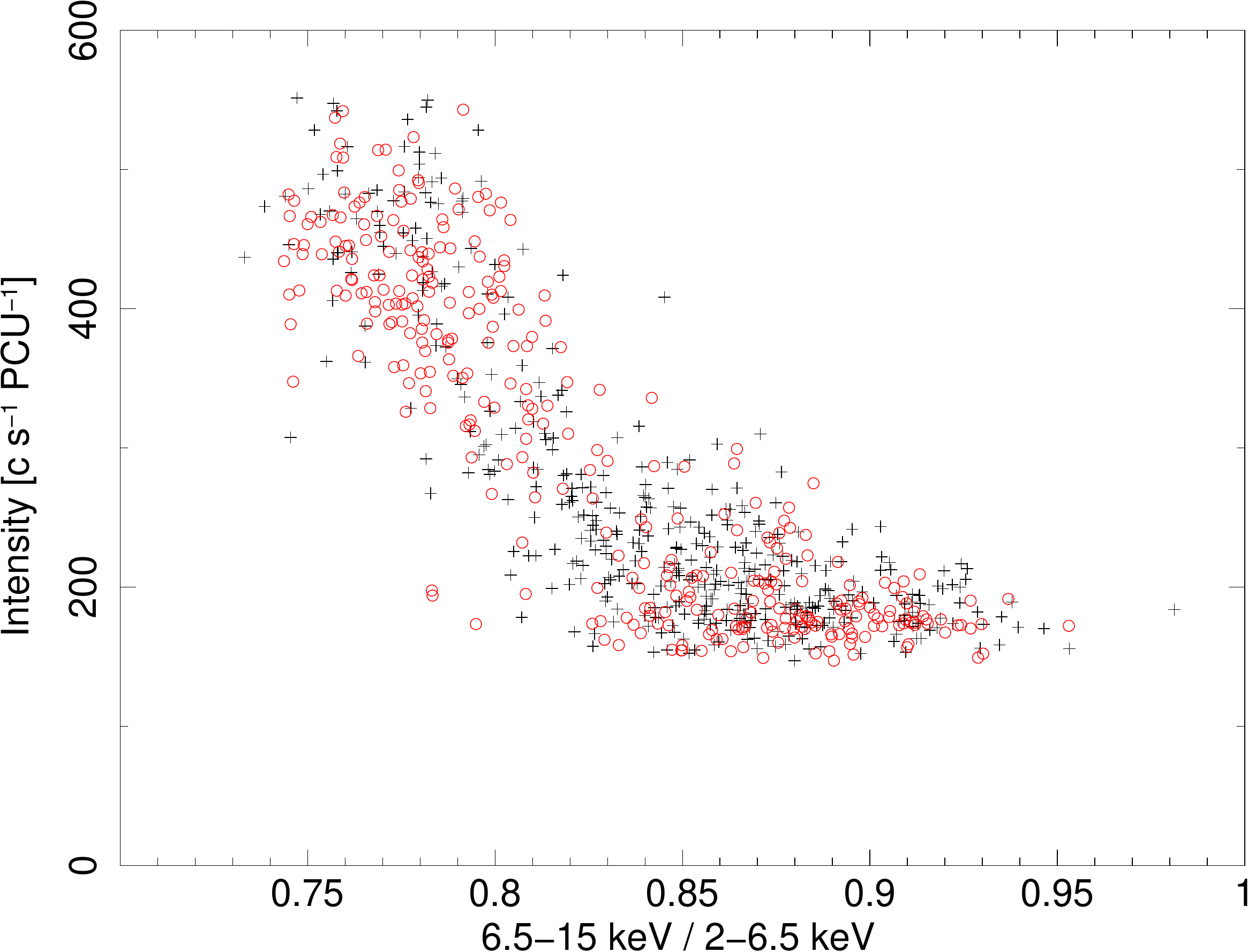}
\includegraphics[width=\columnwidth]{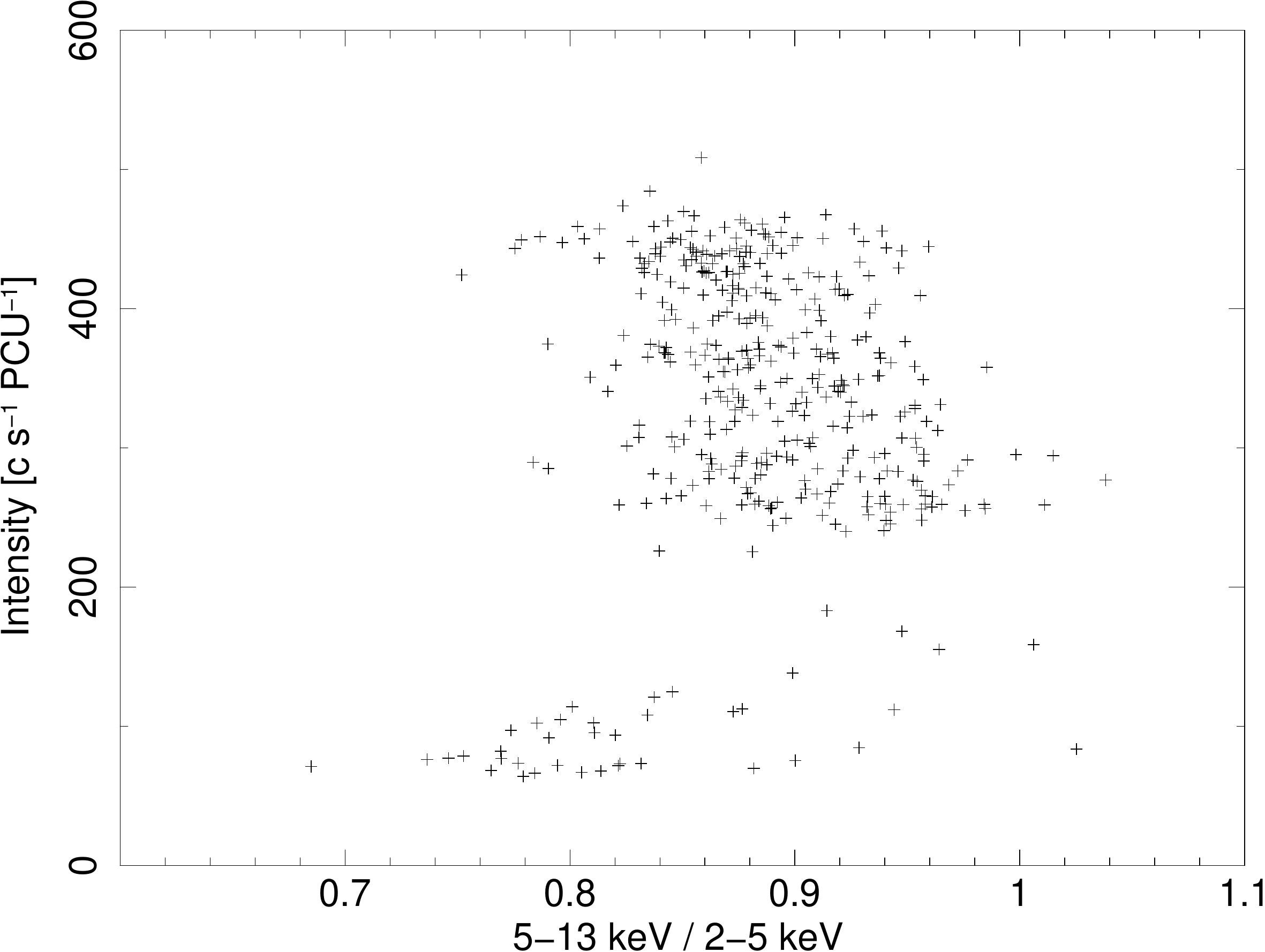}
\caption{Top panel: color-intensity diagram of the first observation at 8-s resolution.
Intensity and color were corrected for the background,
a (changing) collimator response and the contribution by \cont,
assuming that the amplitudes of the last four flares are constant.
The (black) crosses and (red) circles refer
to the rising and decaying portions of the flares.
The color is the same as employed in \citet{2011ApJ...742L..17A}.
Bottom panel: color-intensity diagram for the second observation.
The same corrections were applied as for the top panel,
except for contamination by \cont\ because that source
was outside the field of view for the complete observation.
The color is the same as employed in \citet{2000A&A...355..271B}.
We note the trends are identical for the color employed in the top panel.}
\label{fig:hid}
\end{figure}

In Fig.~\ref{fig:hid}, the intensity versus hardness ratio
diagram is drawn for both observations.
It shows that, in the first observation, the rises and falls of
the oscillations follow the same track,
without hysteresis.

\begin{figure}
\includegraphics[width=\columnwidth]{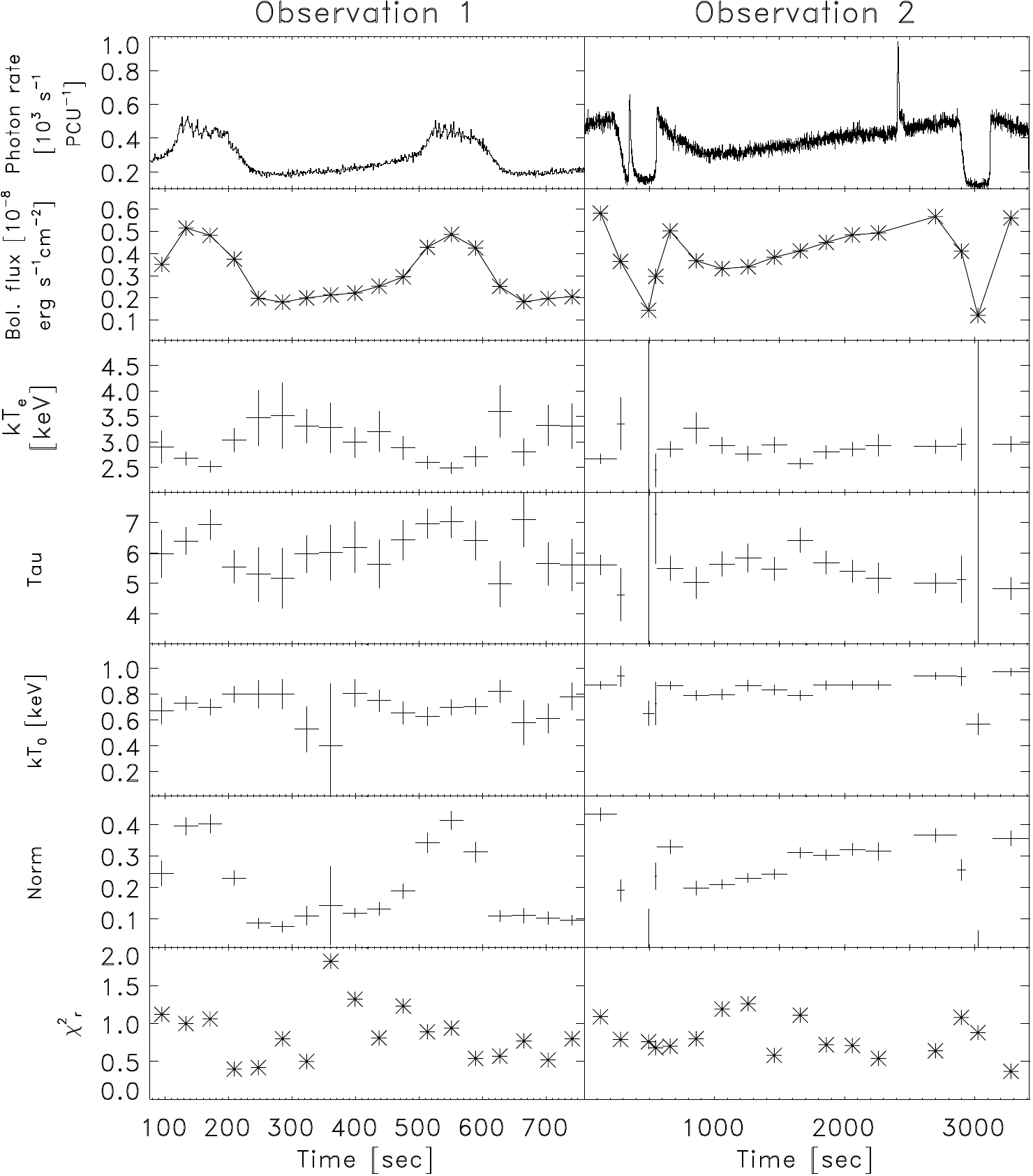}
\caption{Spectral analysis of the two observations, employing a
  Comptonization model. For the first observation, only the uncontaminated
  portion of the data was employed. The entire second observation
  is unaffected by contamination. The type I bursts were ignored.}\label{fig:spectra}
\end{figure}

In general, the spectrum of the RB (including type II bursts, but
excluding type I bursts) is satisfactorily modelled by a Comptonized
spectrum \citep{2015Ba}. Therefore, we applied this model
to the two observations. We only included data for which \cont\
was outside of the field of view. This pertains to the last
0.8~ks of the first observation and the complete second
observation. The results are presented graphically in
Fig.~\ref{fig:spectra}. Indeed, the model is
satisfactory. There is mild spectral variability, with the spectral
hardness anti-correlating with the flux. There is no evidence for
increased absorption during the dips.
The flux change is mostly due to a change in the
normalization, which represents projected emission region size.
The bolometric flux for the applied model is $(2-5)\times$\ergscm{-9}
during the first observation,
and $(1-6)\times$\ergscm{-9} during the second.
Under the same assumptions as in \citet{2015Ba}
this corresponds to 7 -- 18
and 3.5 -- 21 per cent of the Eddington luminosity, respectively.

Contiguous \textit{RXTE} observations span 15 to 3 days before
the Oct. 16, 1999 observation and 3 to 12 days after it,
and 5 to 2 days before and then again 6 days after the Oct. 10, 2008 observation.
In both cases, the source was clearly in the high/soft state in the prior observations,
only emitting type I bursts (for the general spectral and outburst behaviour of the source,
see \citealt{2015Ba}),
while, afterwards,
it had switched to the low/hard state, showing a fainter persistent emission
(by a factor of $\approx$5) and both type I and II bursts.
Both types of behaviour are, therefore,
coincident with the transitional state of the source,
like the $\approx 10$~s dips in some type I bursts in
the same state \citep{2014MNRAS.437.2790B}.
Prior to the second observation, six long ($\sim$60 s)
type I bursts were visible, with a measured recurrence time of 2.2~ks.
The observed $t_{\rm rec}$,
which we have shown to be a tight function of the accretion
rate in the RB \citep{2013MNRAS.431.1947B} is compatible with this
transitional-state identification.

The type I burst and the flaring activity in the first observation
do not seem to affect one another.
Likewise, the two type I bursts in the second observation
show nearly identical peak fluxes and durations,
despite taking place during the dip and well outside it, respectively.
The oscillations and the dips are therefore unaffected by
thermonuclear burning taking place on the surface, and viceversa.

\section{Discussion}
\label{sec:discussion}

We have discovered two new kinds of behaviour
that have never been seen before from the RB,
and add to the complexity of this peculiar source.
Light curves similar to the ones analyzed here
have only been seen in two other LMXBs,
\grs\ and \igr.
Both are thought to harbour black holes (BHs),
although only the former has a dynamical mass measurement
\citep[most recently refined to $12.4^{+2.0}_{-1.8}$~M$_\odot$,][]{2014ApJ...796....2R}.

\grs\ has the largest orbit among LMXBs
\citep[orbital period 33.5~days,][]{2001Natur.414..522G},
has been active for 20 years \citep{1994ApJS...92..469C}
and shows super-Eddington fluxes and superluminal jets 
with a distance of $\approx 12.5$~kpc \citep{1994Natur.371...46M}.
It exhibits at least twelve classes
of variability \citep{2000A&A...355..271B}.
In general, the
extraordinary behavior is attributed to intermittent ejections from
the disk \citep{2002MNRAS.331..745K,2004ARA&A..42..317F}.
\igr\ is less well characterized.
Orbital period, distance and BH mass are unknown.

The first RB observation
highly resembles the $\rho$
(`heartbeat') variability state of \grs\ \citep{2000A&A...355..271B} and
\igr\ \citep{2011ApJ...742L..17A}.
In all three sources, the heartbeats are asymmetric,
decaying faster than they rise.
The $\rho$ behavior is the slowest
in the RB, with timescales of $\approx 2-100$~s in \igr,
and $\approx 40-120$~s for
\grs\ \citep{2011ApJ...737...69N}.
The relative scatter in the timescale of the RB is
similar to that seen in \grs, where the period can vary by
more than 20 per cent between consecutive flares \citep{2011ApJ...737...69N}.
Variability near the maxima of the modulation
is visible in the RB observation and has also been noticed in \grs\
\citep{2000A&A...355..271B}. 
The fractional amplitudes are
similar, 60--75 per cent.

The largest difference in behaviour is that
in \grs\ the peak flux of the modulation is 80--90 per cent
of the Eddington limit \citep{2011ApJ...737...69N}.
In the RB, this is only 18 per cent.
The uncertainty on the luminosity in terms of the
Eddington limit (see \citealt{2015Ba})
cannot bring the two results in agreement.
The spectra behave partially similarly.
The pattern of the RB
in the hardness-intensity diagram (HID, Fig.~\ref{fig:hid})
partly follows the one in \grs\
when the latter is below about 6000~c/s/PCU \citep{2011ApJ...742L..17A},
with hard low-flux intervals and soft high-flux ones.
However, the highest luminosity points reached
by \grs\ are the hardest, and create
a circular hysteresis pattern in the HID
that is not seen in the RB.
Note that \igr\ also shows a hysteresis pattern,
although in an opposite sense to \grs.

The intrinsic luminosity of \igr\ is not known,
due to the lack of a distance measurement.
Assuming the source emitted at the Eddington limit like \grs,
this would imply either a distance in excess of 20~kpc,
with a galactic latitude suspiciously high above the disk,
or an extremely low mass BH
\citep[$<3 M_{\odot}$, ][]{2011ApJ...742L..17A,2012ApJ...757L..12R}.
However, hydro-dynamical simulations by \citet{2014arXiv1411.4434J}
found that a thermal-viscous instability
led by radiation-pressure domination in the innermost region of disc
can occur for luminosities as small as 10 per cent of the Eddington limit,
and can reproduce the time scales and amplitudes of the heartbeat light curves.
The disc is stable at smaller luminosities,
while larger ones produce, in regions of the disc further out,
a wind so strong it effectively depletes the inner disc,
killing off the instability.
Their hypothesis is supported by the detection of
strong absorption lines (due to outflows) in the spectrum of \igr,
which become undetectable when
the luminosity decreases and
the instability becomes visible \citep{2012ApJ...746L..20K}.
A sub-Eddington luminosity
would bring \igr\ better into agreement with the
radio/X-ray correlation for BH \citep{2011A&A...533L...4R}.
Clearly, the appearance of the $\rho$ class in the RB
at roughly the same fraction of the Eddington limit supports this scenario.

A possible explanation for the lack of hysteresis
in the RB as shown by the BH accretors during the $\rho$ variability
could be related to the presence of an additional
emission component from the solid surface
and/or boundary layer of the former.
As shown by \citet{2003MNRAS.342.1041D},
NS and BH accretors clearly show different
spectral evolutions as a function of luminosity in their bright states,
which is due to the presence of a surface in the case of NSs.

In all three sources, the flares in the heartbeat are asymmetric: they
decay sharply, to then rise more gradually.
This is true for all hard phases in \grs\,
which \citet{1997ApJ...488L.109B,1997ApJ...479L.145B}
interpreted as the result of a viscous-thermal instability in the
accretion disk, with the rise time determined
by the speed at which a heating wave moves through the disc,
and the faster decay time corresponding
to the infall of matter into the BH.

The second observation highly resembles the $\theta$-class
variability of \grs\ (not seen in \igr).
This class also presents ``M''-shaped light curves
separated by 100-200~s long minima.
The luminosities are again lower in the RB than in \grs\,
where this variability reaches the Eddington limit
\citep{2000A&A...355..271B}.
The spectral behaviour of the RB is similar
to that observed in \grs,
in which the ``M'' portions are harder,
while the dips are slightly softer.
The separation is however not as clear as in \grs.

A particularly interesting feature of the $\theta$ variability
in the RB is the occurrence of an unaffected type I burst during
the dip.
The burst reveals that the NS emission must be unaffected.
Still, the accretion disk flux drops by
75 per cent within a matter of 60~s.
This points either to a cylindrically
asymmetric disk which is obscured from below
or above our line of sight to the NS
or to a global change of the emissivity of the disk,
for instance by a loss of mass.
The latter would be consistent with the ideas about \grs.

Just as there are similarities between the RB and \grs,
there are similarities between the RB the BP:
like the RB, the BP shows type II bursts.
\grs\ and the BP are
physically better characterized than the RB and it is tempting to
search for commonalities and attribute those to the RB as well.
Both sources have relativily long orbital periods of 33.5 and 11.8 d,
respectively.
\igr\ may also have a large orbital period
\citep{2012MNRAS.422L..91W,2014arXiv1406.7262G}.
A long orbital period points to
a relatively large accretion disk, which naturally explains the large
accretion rates. It may be smaller for the RB, because its transient
outbursts last shorter and recur more frequently, but still longer
with respect to the average LMXB.
\citet{2010ApJ...718..620W} have shown that a single relation
exists between the maximum outburst luminosity as a fraction of Eddington
and the orbital period for NS and BH LMXBs.
Substituting the maximum observed persistent luminosity $F\sub{peak} = 0.45 F\sub{Edd}$
\citep{2013MNRAS.431.1947B} yields indeed a relatively large
$P\sub{orb} = 7.8$~days for the RB.

It is striking that similar behaviour is present between the RB,
a NS system, and the two BH systems, even more because
they are all peculiar sources among their own classes.
Also, both the observations analyzed here take place
at the transition between a type-II-burst-free state
and an active type II bursting phase, which suggests
that type II bursts and these variabilities might be connected.
With respect to this, we note that apart from
variability classes $\theta$ and $\rho$,
\grs\ shows light curves
\citep[particularly those of variability classes $\mu$ and $\lambda$, see][]{2000A&A...355..271B}
that resemble the type II burst behavior in the RB.

However, there are important differences.
The type II bursts are well known to show very little
hardness variations
\citep[see e.g., Appendix~B in ][]{2015Ba},
whereas the intensity variations during variability in \grs\
always correspond to spectral transitions.
Also, while the type II burst rise times are always shorter
than the decay time
(which is why they can be mistaken for type I bursts and viceversa)
and their duration determines the waiting time to the next burst,
the opposite is true for the $\lambda$ variability class in \grs,
where rises are slower than decays
and the duration of a burst correlates with that of the previous quiescent phase,
compatibly the thermal instability picture mentioned above
\citep{1997ApJ...488L.109B}.
Finally, type II bursts in the RB only appear
below a critical persistent luminosity of about $0.1 L\sub{Edd}$
\citep{2015Ba}, whereas states of \grs\ with no large amplitude variations
($\phi$ and $\chi$) are not clearly separated in luminosity from the other ones.

Nonetheless, these differences might be related to the nature of the accretor,
for example because NSs possess a strong magnetic dipole field.
It would be interesting to perform complete
population studies of the X-ray bursts
in the BH systems and the BP,
to carry out a comparison with the results by \citet{2015Ba} for the RB.

\section*{Acknowledgements}
We would like to thank Jeroen Homan, Peter Jonker, Michiel van der Klis,
Alessandro Patruno and Manuel Torres
for the useful discussion and commentary provided,
and the anonymous referee for their suggestions.

\footnotesize{
  \bibliographystyle{mn2e}
  \bibliography{paper}
}

\label{lastpage}
\end{document}